\documentclass[apjl]{emulateapj}
\usepackage{amssymb}
\usepackage{amsbsy}
\usepackage{color}
\usepackage{mathrsfs}
\def \mathbi#1{\textbf{\em #1}}
\newcommand\pasa{{Publ.~Astron.~Soc.~Aust.}}%

\slugcomment{ApJ Letters, accepted}

\begin{document}

\shorttitle{How black holes accrete}
\title{Tearing up the disc: how black holes accrete}
\shortauthors{\sc{C.~J.~Nixon, A.~R.~King, D.~J.~Price \& J.~Frank}}
\author{Chris~Nixon\altaffilmark{1}, Andrew~King\altaffilmark{1}, Daniel~Price\altaffilmark{2}, Juhan~Frank\altaffilmark{3}} 
\altaffiltext{1}{Department of Physics
  \& Astronomy, University of Leicester, Leicester LE1 7RH UK}
\altaffiltext{2}{Monash Centre for Astrophysics (MoCA), School of
  Mathematical Sciences, Monash University, Vic.  3800, Australia}
\altaffiltext{3}{Department of Physics and Astronomy, Louisiana State
  University, Baton Rouge, LA 70803-4001, United States}
\email{chris.nixon@leicester.ac.uk}

\begin{abstract}
We show that in realistic cases of accretion in active galactic nuclei or
stellar--mass X--ray binaries, the Lense--Thirring effect breaks the central
regions of tilted accretion discs around spinning black holes into a set of
distinct planes with only tenuous flows connecting them.  If the original
misalignment of the outer disc to the spin axis of the hole is $45^{\circ}
\lesssim \theta \lesssim 135^{\circ}$, as in $\sim 70$\% of randomly oriented
accretion events, the continued precession of these discs sets up partially
counter--rotating gas flows. This drives rapid infall as angular momentum is
cancelled and gas attempts to circularize at smaller radii. Disc breaking
close to the black hole leads to direct dynamical accretion, while breaking
further out can drive gas down to scales where it can accrete rapidly. For
smaller tilt angles breaking can still occur, and may lead to other observable
phenomena such as QPOs. For such effects not to appear, the black hole spin
must in practice be negligibly small, or be almost precisely aligned with the
disc. Qualitatively similar results hold for any accretion disc subject to a
forced differential precession, such as an external disc around a misaligned
black hole binary.
\end{abstract}

\keywords{accretion, accretion disks --- black hole physics --- hydrodynamics
  --- galaxies: active --- stars: neutron}

\section{Introduction}
\label{intro}
Accretion discs are common in astrophysics on all scales from protostars to
AGN \citep[see e.g.][]{Pringle1981,Franketal2002}. Many treatments assume that
the disc is aligned with the symmetry axis of the central object, although
there is often no {\it a priori} reason for this. The first widely--studied
case relaxing this restriction was the evolution of tilted discs around
spinning black holes. Until recently the standard picture of tilted disc
evolution was that, in the regime where viscosity acts diffusively
(technically, $\alpha > H/R$), the inner disc would align or counteralign
rapidly with the hole's spin, with a smooth warp to the still misaligned outer
parts. This is often called the Bardeen--Petterson effect \citep[][but note
  their equations do not conserve angular momentum; see
  \citealt{PP1983,Pringle1992,Ogilvie1999} for detailed discussions of the
  correct equations; and \citealt{Kingetal2005} for the possibility of
  counteralignment]{BP1975}.

In a recent paper, \citet{NK2012} showed that this evolution can be very
different for large inclinations of the disc and spin, and/or low values of
the dimensionless viscosity coefficient $\alpha$ \citep{SS1973}. Enforcing the
connections imposed by conservation laws between the various components
(`radial', and `vertical') of viscosity \citep{Ogilvie1999}, \citet{NK2012}
showed that the viscous torques in the disc may be unable to communicate the
Lense--Thirring precession efficiently enough to produce a smooth
warp. Instead one expects a sharp break in the disc plane between the aligned
inner parts and the misaligned outer parts, connected only by tenuous rings of
gas with inclinations changing rapidly across the break. We shall see from our
Eqn.~\ref{crit} below, that even if the viscosity coefficients remained
constant with warp amplitude, the disc would still break for realistic
parameters. \cite{LP2010} show that an assumed break of this type remains
stable in 3D simulations of such discs. \cite{Nixonetal2012} show that
inclined, partially counter--rotating gas orbits within an accretion disc lead
to cancellation of angular momentum and thus subsequent accretion. The mass
flow rate through the disc can, for a time, be increased by large factors up
to $\sim 10^4$ times that of the corresponding disc with zero inclination.

These results suggest that sufficiently inclined discs might break, and that
if the precession rate of the inner and outer disc differs enough, discs
rotating in opposed senses might interact and produce dynamical mass
infall. We consider these questions in this Letter. We ask\\
1) for realistic parameters, can the Lense--Thirring 
torques exceed the local viscous torques in the disc?\\
2) If they can, can we arrange the broken disc such that disc orbits
counter--rotate?

\section{Where does the disc break?}
\label{where}

Here we estimate analytically the radius at which the disc is likely
to break. This will give us an idea of the parameters which lead to
breaking, and whether it is a common event. To break the disc, the
torque resulting from the Lense--Thirring effect must overcome the
local viscous torques, or equivalently, the orbits in the disc must
precess faster than the viscosity can communicate the precession. To
illustrate this point, let us imagine the two extremes. If the
viscosity is dominant, the precession is communicated throughout the
disc instantaneously, and the whole disc precesses rigidly. At the other
extreme where viscosity is negligible, orbits at different radii
precess at different rates and the disc must break into many distinct
rings. The nonlinear connection between effective viscosity coefficients,
enforced by conservation laws \citep{Ogilvie1999}, tell us that once a
disc starts to break in this way, the viscosity evolves so as to
reinforce the tendency to break \citep{NK2012}. 

To calculate the breaking radius for a given viscosity we can assume that the
disc has no initial warp. Then we can consider the usual viscous torque, and
to a good approximation neglect the more complicated physics of a warped
disc. The azimuthal viscous force per unit area in the disc is
proportional to the rate of shear $R {\rm d}\Omega/{\rm d}R$ (where $R$ is the
radial coordinate and $\Omega$ is the disc angular velocity) and so can be
written as
\begin{equation}
f_{\nu} = \mu R\frac{{\rm d}\Omega}{{\rm d}R}.
\end{equation}
where $\mu$ is the dynamical viscosity.  The area of an interface in
the disc is $2\pi R H$, where $H$ is the disc vertical
thickness. So the viscous force acting in the azimuthal
direction is given by
\begin{equation}
F_{\nu} = 2\pi R H \mu R\frac{{\rm d}\Omega}{{\rm d}R} = 2\pi R \nu
\Sigma R \frac{{\rm d}\Omega}{{\rm d}R},
\end{equation}
where we have substituted the dynamic viscosity for the kinematic
viscosity $\left(\mu = \rho\nu\right)$, and used $\Sigma = \rho
H$. The magnitude of the viscous torque $G_{\nu}$ is given by
\begin{equation}
G_{\nu} = \left|\mathbi{R}\times\mathbi{F}_{\nu} \right| = 2\pi R \nu
\Sigma R^2 \left(-\Omega^\prime\right),
\end{equation} 
where the prime denotes the radial derivative. This is the usual viscous
torque at the interface of two annuli in an accretion disc
\citep[see][]{LP1974,Franketal2002}. For near--Keplerian rotation ($\Omega^2
\approx GM/R^3$) it becomes
\begin{equation}
\label{nutorque}
G_\nu = 3\pi\nu\Sigma\left(GMR\right)^{1/2}.
\label{G+}
\end{equation}

The Lense--Thirring precession induces a torque with magnitude
\begin{equation}
\label{LTtorque}
G_{\rm LT} = 2\pi R H \left|\boldsymbol{\Omega}_{\rm p} \times
\mathbi{L}\right| = 2\pi R H \Omega_{\rm p}\Sigma R^2
\Omega\left|\sin\theta\right|,
\end{equation}
where the Lense--Thirring frequency $\boldsymbol{\Omega_{\rm p}} = 2G\mathbi{J}_{\rm h}/c^2R^3$, the disc angular momentum density
$\left|\mathbi{L}\right| = \Sigma R^2 \Omega$ and $\theta$ is the angle between
the angular momentum of the black hole and the disc. We assume the disc breaks
when the Lense--Thirring torque tears gas off the disc faster than viscosity
can make it spiral inwards, which requires
\begin{equation}
G_{\rm LT} \gtrsim G_\nu.
\end{equation}
Making the standard assumption of a thin Keplerian $\alpha$--disc
\citep{SS1973} and using $J_{\rm h} = aGM^2/c$, the radius at which we expect the disc to break is given by
\begin{equation}
R_{\rm break}\lesssim
\left(\frac{4}{3}\left|\sin\theta\right|\frac{a}{\alpha}\frac{R}{H}\right)^{2/3} R_{\rm g},
\label{crit}
\end{equation}
where $R_g = GM/c^2$ is the gravitational radius of the black hole.

Equation (\ref{crit}) looks plausible; it makes sense that at large
viscosity, low spin or small inclination angles the disc cannot be
broken, i.e. $R_{\rm break} < R_{\rm g}$. Conversely, for low
viscosity, high spin and/or large inclinations we expect the disc to
break into distinct rings at some radius $R_{\rm break} > R_g$.

The typical radius at which the disc breaks is given by
\begin{equation}
R_{\rm break} \lesssim 350 R_{\rm g}
\left|\sin\theta\right|^{2/3}\left(\frac{a}{0.5}\right)^{2/3}
\left(\frac{\alpha}{0.1}\right)^{-2/3}\left(\frac{H/R}{10^{-3}}\right)^{-2/3}
\end{equation}
where we have parameterized using quantities typical for AGN discs. This
radius falls within typical discs, suggesting that inclined discs near
spinning black holes are quite susceptible to breaking. We caution that at
extreme parameters this simple argument may not suffice to predict the
behaviour of the system, although we expect the general behaviour to hold. In
the next section we confirm the breaking of the disc with numerical
simulations. These simulations are a preliminary investigation into this
problem and we intend to follow up in more detail in future publications.

\section{Counter--rotation}
\label{geometry}
Let us consider an inclined disc that breaks under the action of a strong
differential precession. Its inner and outer regions precess almost
independently at different rates.  The precession timescale is much shorter
than the alignment timescale, which must wait for precession to induce
dissipation. So both the inner and outer discs retain their inclinations to
the black hole spin. The outer disc remains almost unmoved, while the inner
disc (typically a ring of radial width $\sim H$) precesses rapidly about the
spin axis. If the angle $\theta$ between the outer disc and the hole spin lies
between $\sim 45 - 135^{\circ}$ the inner ring must form an angle $2\theta >
90^\circ$ with respect to the outer disc after half a precession period. The
rotational velocities are now partially opposed.
 
This configuration is similar to those adopted in the counter--rotating disc
simulations in \cite{Nixonetal2012} and so must result in rapid accretion. We
note the probability that a randomly oriented accretion event lies in the
critical range of inclinations is given by the fractional solid angle as
$\cos\left(\pi/4\right)$ i.e. $\approx 70$\%. In other words, disc breaking
and dynamical infall from counter--rotating accretion flows must be common in
active galactic nuclei. It is also common in stellar--mass
X--ray binaries if the spin of the black hole (or neutron star) accretor is
sufficiently misaligned with the binary plane.

We report two simulations of an inclined disc around a spinning black hole
using the Smoothed Particle Hydrodynamics (SPH) code \textsc{phantom}
\citep[see e.g.][]{PF2010,LP2010,Nixonetal2012,Nixon2012}. SPH performs well
in modelling warped discs \citep{LP2010} finding excellent agreement with the
analytical treatment of \cite{Ogilvie1999}. This is to be expected as both
treatments solve the Navier Stokes equations with an isotropic viscosity. The
connections between the viscosity coefficients derived by \cite{Ogilvie1999}
therefore naturally holds in our numerical treatment. In the simulations
reported here we implemented the Lense--Thirring effect, following
\cite{NP2000}. The simulations use a disc viscosity with Shakura \& Sunyaev
$\alpha \simeq 0.1$ \citep[cf.][]{LP2010}, a disc angular semi--thickness of
$H/R \simeq 0.01$ and the black hole has a spin $a=1$. Initially the disc has
no warp and extends from an inner radius of $50R_{\rm g}$ to an outer radius
of $250R_{\rm g}$, with a surface density profile $\Sigma = \Sigma_0
(R/R_0)^{-p}$ and locally isothermal sound speed profile $c_{\rm s} = c_{{\rm
    s},0} (R/R_0)^{-q}$ where we have chosen $p=3/2$ and $q=3/4$ to achieve a
uniformly resolved disc \citep{LP2007}. The disc is initially composed of 2
million particles, which for this setup gives $\left<h\right>/H \approx 0.8$
\citep[cf.][]{LP2010}. The two simulations differ only by the relative
inclination angle to the black hole.

\begin{figure}
  \begin{center}
    \includegraphics[angle=0,width=\columnwidth]{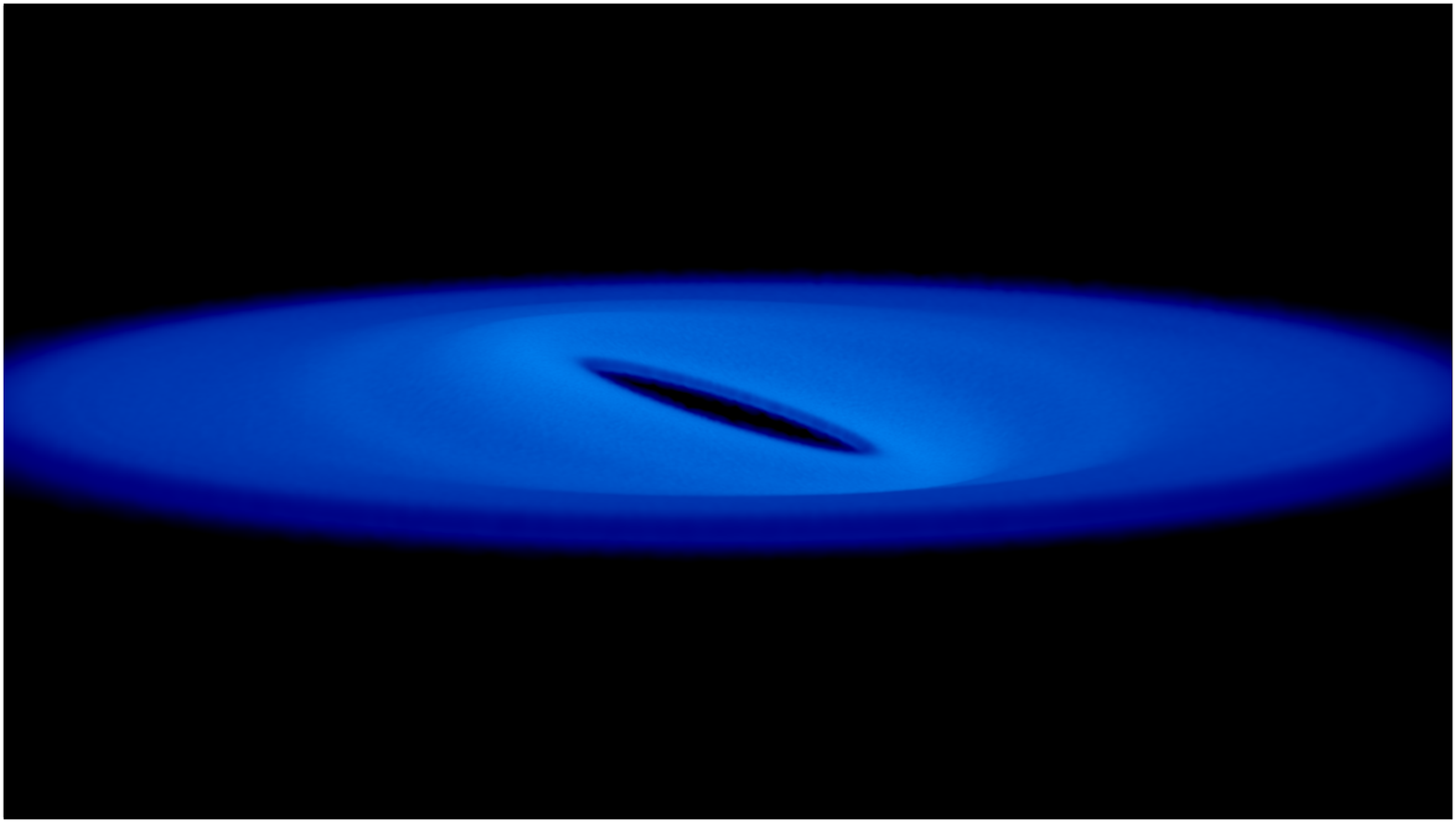}
    \caption{Full 3D surface rendering of the small inclination
      simulation. The whole disc was initially inclined at $10^{\circ}$ to the
      hole with no warp. This snapshot is after approximately 500 dynamical
      times at the inner edge of the disc ($50 R_{\rm g}$).}
    \label{small}
  \end{center}
\end{figure}

\begin{figure}
  \begin{center}
    \includegraphics[angle=0,width=\columnwidth]{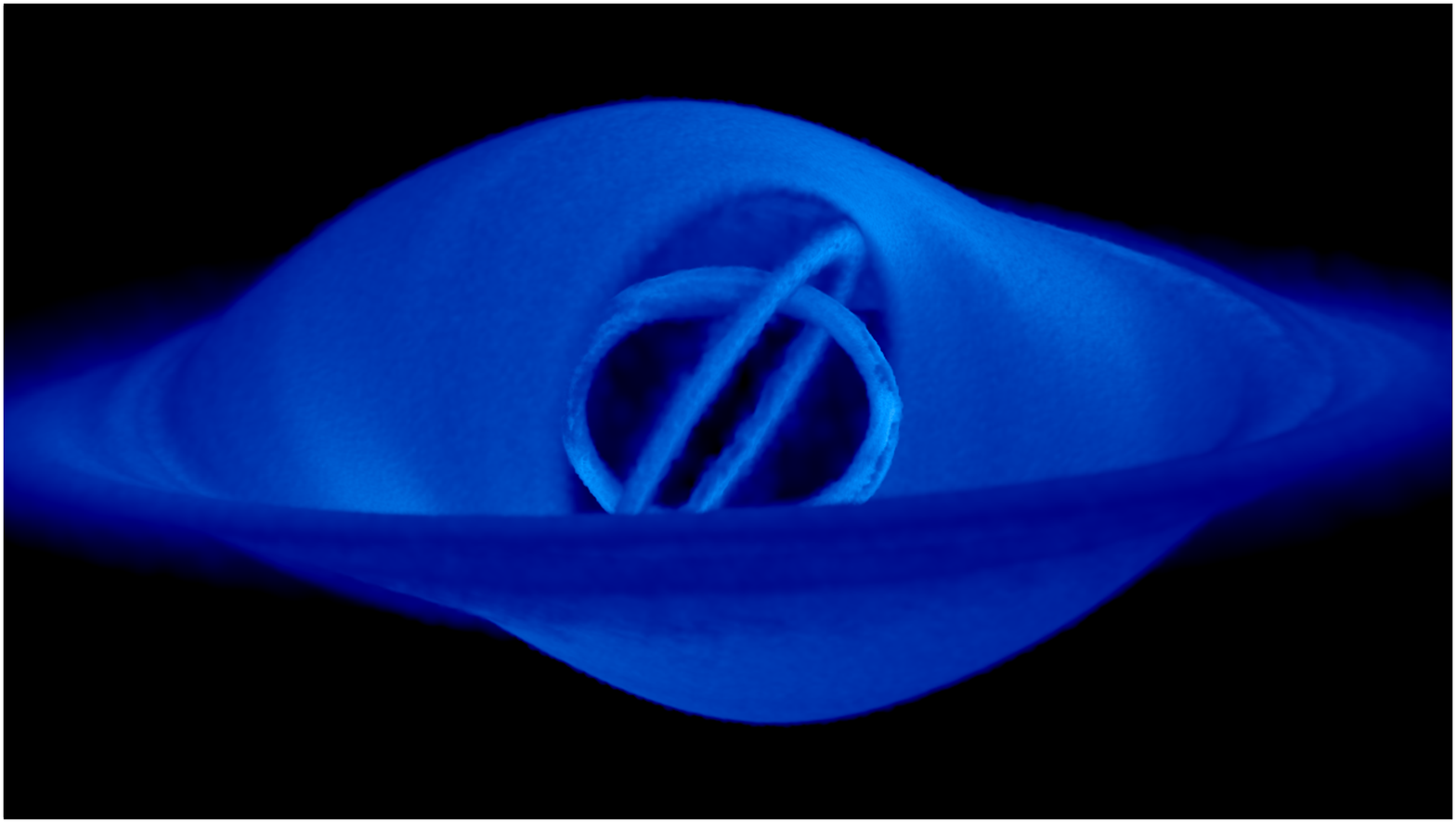}
    \caption{Full 3D surface rendering of the large inclination
      simulation. The whole disc was initially inclined at $60^{\circ}$ to the
      hole with no warp. This snapshot is after approximately 500 dynamical
      times at the inner edge of the disc ($50 R_{\rm g}$).}
    \label{large}
  \end{center}
\end{figure}

\begin{figure}
  \begin{center}
    \includegraphics[angle=0,width=\columnwidth]{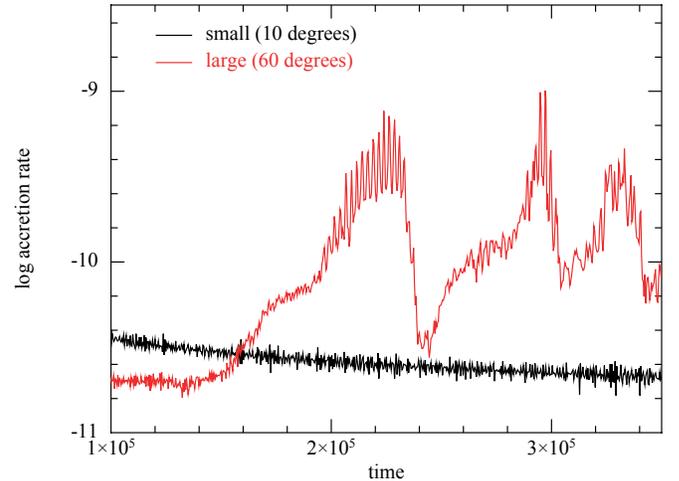}
    \caption{The accretion rates with time for the small (black) and
      large (red) inclination simulations. The accretion rate is in
      arbitrary units. The time unit is the dynamical time at
      $R=R_{\rm g}$, i.e. the dynamical time at the inner edge of the
      discs is $\sim 350$. Note this is the mass flow rate through
      $R=50 R_{\rm g}$ and not the final accretion rate on to the
      black hole.}
    \label{accretion}
  \end{center}
\end{figure}

The simulation shown in Fig.~\ref{small} has an initial inclination of
$10^\circ$, and thus $R_{\rm break} \approx 40 R_{\rm g}$, i.e. at a radius
inside the disc's inner boundary, and so we do not expect the disc to
break. This agrees with the simulation, which shows the usual
(Bardeen--Petterson) evolution with a smooth warp. In contrast, the simulation
shown in Fig.~\ref{large} has an inclination of $60^\circ$ and therefore
$R_{\rm break} = 110 R_{\rm g}$, i.e. we expect the disc to break. The
simulated disc does indeed break, producing multiple distinct rings of gas
with large relative inclinations. This leads to phases of strong accretion
when the rings are highly inclined, and quieter phases when they are not\footnote{Movies of the simulations in this paper are available at http://www.astro.le.ac.uk/users/cjn12/tearing.shtml}. The
accretion rates for the two cases are shown in Fig.~\ref{accretion}.


We note that this evolution is more extreme than the 1D simulations
reported in \citet{NK2012}. However the numerical method used there assumed
the gas always remained on circular orbits, evolving purely by viscous
diffusion. This was appropriate for the problem studied there, namely, whether
such a relatively orderly disc could break at all, given the nonlinear
evolution of the viscosity predicted by \cite{Ogilvie1999}. The current paper
studies a dynamical problem, where the inclination of disc orbits can change
so rapidly that viscous diffusion of gas in circular rings is no longer an
adequate approximation. Interestingly, \cite{NK2012} did remark that their
simulated disc rings appeared to be trying to break into more than one
distinct plane.

\section{Discussion}
\label{discussion}
We first discuss the possible arguments against this picture. The main unknown
in this work is the nature of the viscosity controlling angular momentum
transport in the disc. In this paper we have assumed that this can be modelled
as an isotropic Shakura \& Sunyaev $\alpha$ viscosity. There is a strong basis
for assuming that the radial transport of angular momentum (governed by the
azimuthal viscosity) is limited to $\alpha \sim 0.1$ \citep{Kingetal2007} and
we expect discs around black holes to be very thin away from the immediate
vicinity of the strongly accreting hole \citep[e.g.][and references
  therein]{KP2007}. However, the nature of the viscosity is unknown. In
reality the local viscosity is likely to result from MHD effects
\citep{BH1991}, and may well be anisotropic. The azimuthal shear is likely to
be secular, with gas parcels continually moving away from each other, whereas
the vertical shear is probably oscillatory \citep{Pringle1992}. This is
suggestive of a favourable anisotropy where the vertical viscosity does not
strongly oppose breaking, but the result is simply not known. The consistency
requirements worked out by \cite{Ogilvie1999} for a locally isotropic
viscosity show that in a strong warp, the viscosity trying to hold the disc
together is likely to weaken. There appears no reason to suggest this differs
for an anisotropic viscosity.

Another possible complication is the thermal evolution of the gas. As the disc
orbits do not all lie in the equatorial plane, they must shock and heat up. In
the simulations above we have assumed an isothermal equation of state, so this
extra heat is assumed to be radiated away instantly. This is reasonable, as
the densities in black hole discs are high and cooling is likely to be
efficient. However if the disc cannot cool on the local precession timescale,
it may heat up significantly so that Eq.~\ref{crit} is no longer satisfied. In
this case the disc may rapidly thicken, perhaps even becoming thermally
unstable, and switch to a different mode of both accretion and warp
propagation.  On the other hand, \citet{Nixonetal2012} found, in
counter--rotating disc simulations using an adiabatic equation of state, that
although the gas dynamics can be strongly modified by gas heating, the net
result in terms of rapid accretion is similar.

Otherwise there appears to be no obvious reason why this behaviour should be
suppressed. We therefore expect this to be generic to most cases of accretion
on to black holes (or neutron stars, since the Lense--Thirring effect applies
here too), and more generally for the evolution of gas discs in the presence
of a strong precession.

\section{Conclusions}
\label{conclusions}
We have shown that for realistic parameters a randomly oriented
accretion event on to a spinning black hole is likely to form a disc
which is susceptible to breaking at a radius close to the hole. If the
angle between the disc and the hole lies between $\sim 45 - 135^{\circ}$
the interaction of partially opposed gas motions is likely, and
leads to cancellation of angular momentum and rapid infall.

This quasi--dynamical form of accretion, which appears to be a generic
consequence of randomly oriented accretion on to a black hole,
significantly alters the standard picture of slow viscous
accretion. There is nothing to prevent a succession of events where
rings break off the inner disc edge and then precess independently of
those inside or further out (see Fig.~\ref{large}).  It is reasonable
to think of this process as tearing up the disc in a chaotic way.

We shall consider possible consequences of this picture in subsequent
papers, but note several points here.

1. Tearing the disc can lead to rapid gas infall, but the long--term
rate of central accretion is ultimately controlled by the outer disc.

2. In a stellar--mass binary system this means that tearing {\it
  modulates} a quasi--steady mass transfer rate. The modulation might
have large amplitude if the disc/spin inclination is high. Even if the
inclination is modest, there are likely to be observable effects which
could include quasiperiodic behaviour such as quasiperiodic
oscillations (QPOs). Still more effects can occur if the infalling
rings shadow the central X--ray source.

3. By contrast, in active galactic nuclei, the viscous timescale of
the outer disc may easily exceed a Hubble time, and no steady state is
ever set up. Thus tearing of a significantly inclined AGN disc may
promote significant accretion when the central black hole would
otherwise not gain mass at all.

4. The torques we have considered here are all internal to the disc --
black hole system. They cannot affect any conclusions concerning the
global conservation of angular momentum or mass of this system. In
particular, considerations of the long--term evolution of black hole
spin through accretion remain unchanged, whether the accretion is
assumed to be coherent \citep[e.g.][]{VR2005,BV2008} or chaotic
\citep{KP2006,KP2007,Kingetal2008}.

5. We can expect qualitatively (and sometimes quantitatively)
similar effects for other cases where an accretion disc is subject
to a forced external differential precession. Most obviously,
\cite{Nixonetal2011b} have shown that the effective potential
experienced by a disc accreting on to a misaligned binary, as is
thought to occur when supermassive black holes are close to
coalescence, is extremely similar to that caused by the
Lense--Thirring effect. For initial inclinations near to co-- or
counter--rotation the disc respectively coaligns or counteraligns
\citep[for the subsequent evolution of prograde circumbinary discs
  see][and for retrograde circumbinary discs see
  \citealt{Nixonetal2011a}]{Cuadraetal2009,Lodatoetal2009}. However,
disc tearing changes this picture and may well bring gas into the
close vicinity of the holes on near--dynamical timescales, and thus
help with the last parsec problem \citep{Begelmanetal1980} as well as
feeding problems.

6. In general any black hole has some spin, and in general any
accretion disc plane may be inclined to this spin. To prevent any of
the effects we have discussed here from appearing, the misalignment
must satisfy
\begin{equation}
|\sin\theta| \la {3\alpha\over 4a}{H\over R},
\end{equation}
which is extremely small for realistic parameters. For example, a moderately
thick disc with $H/R=0.1$, $\alpha=0.1$ and a low spin $a=0.1$ must be
inclined by less than $4^\circ$ to avoid this process. We suggest that disc
tearing, particularly in the inner disc, probably occurs in many if not most
cases of black hole or non-magnetic neutron star accretion.


\acknowledgements We thank Jim Pringle and Yuri Levin for useful discussions
and the referee for a very helpful report. We used \textsc{splash}
\citep{Price2007} for the visualisation. Research in theoretical astrophysics
at Leicester is supported by an STFC Rolling Grant. CJN was supported by an
STFC research studentship during part of this work. This research used the
ALICE High Performance Computing Facility at the University of Leicester. Some
resources on ALICE form part of the DiRAC Facility jointly funded by STFC and
the Large Facilities Capital Fund of BIS.

\bibliographystyle{apj} 

\end{document}